\documentclass[epj,final]{svjour}

\usepackage{graphicx}
\usepackage{dcolumn}   
\usepackage{bm}        
\usepackage{amssymb}   
\usepackage{textcomp}
\usepackage{ucs}
\usepackage{hyperref}
\usepackage{amsmath}
\usepackage{color}
\usepackage{footnote}

\begin{document}

\title{Intense Vacuum-Ultraviolet and Infrared Scintillation of Liquid Ar-Xe Mixtures}

\author{A. Neumeier\inst{1} \and T. Dandl\inst{2} \and T. Heindl\inst{2} \and A. Himpsl\inst{2} \and L. Oberauer\inst{1} \and W. Potzel\inst{1} \and S. Roth\inst{1} \and S. Sch\"onert\inst{1} \and \\J. Wieser\inst{3} \and A. Ulrich\inst{2}\thanks{\emph{Andreas Ulrich:} andreas.ulrich@ph.tum.de 
}
}                     
\institute{Technische Universit\"at M\"unchen, Physik-Department E15, James-Franck-Str. 1, D-85748 Garching, Germany \and Technische Universit\"at M\"unchen, Physik-Department E12, James-Franck-Str. 1, D-85748 Garching, Germany \and excitech GmbH, Branterei 33, 26419 Schortens}

\date{Published in EPL (2015)}

\abstract{
Vacuum ultraviolet light emission from xenon-doped liquid argon is described in the context of liquid noble gas particle detectors. Xenon concentrations in liquid argon from 0.1\,ppm to 1000\,ppm were studied. The energy transfer from the second excimer continuum of argon ($\sim$\,127\,nm) to the second excimer continuum of xenon ($\sim$ 174\,nm) is observed by recording optical emission spectra. The transfer almost saturates at a xenon concentration of $\sim$10\,ppm for which, in addition, an intense emission in the infrared at a peak wavelength of 1.17\,$\mu m$ with $13000 \pm 4000$ photons per MeV deposited by electrons had been found. The corresponding value for the VUV emission at a peak wavelength of 174\,nm (second excimer continuum of xenon) is determined to be $20000 \pm 6000$ photons per MeV electron energy deposited. Under these excitation conditions pure liquid argon emits $22000 \pm 3000$ photons per MeV electron energy deposited at a peak wavelength of 127\,nm. An electron-beam induced emission spectrum for the 10\,ppm Ar-Xe liquid mixture ranging from 115\,nm to 3.5\,\,$\mu m$ is presented. VUV emission spectra from xenon-doped liquid argon with exponentially varied xenon concentrations from 0.1\,ppm to 1000\,ppm are also shown. Time structure measurements of the light emissions at well-defined wavelength positions in the vacuum ultraviolet as well as in the near-infrared are presented.
}           

\maketitle

\section{Motivation}

Liquid argon is frequently used or considered as scintillating material in various experiments in rare event physics, e.g., direct dark matter search: ArDM \cite{ArDM}, DarkSide \cite{DarkSide}, MiniCLEAN \cite{MiniCLEAN}, DEAP-3600 \cite{DEAP_1,DEAP_2}, DARWIN \cite{DARWIN} or high energy neutrino physics: ICARUS \cite{ICARUS}, LBNE \cite{LBNE}. In this context it had been found that a small admixture of xenon to the liquid can improve the scintillation efficiency \cite{Pollmann}. 

Excitation of liquid or dense noble gases leads to the emission of the so-called second excimer continuum in the vacuum ultraviolet (VUV) spectral range \cite{Excimer_Buch}. It is known from the gas phase that adding xenon to argon leads to a strong modification of the emission spectrum \cite{Efthimiopoulos}. The emission is shifted from the second continuum of argon at 126\,nm to the second continuum of xenon at 172\,nm. This shift occurs already at small admixtures of xenon of $\sim$1\,\% in the gas phase. A series of argon-gas spectra with a systematic variation of xenon content can be found in \cite{Efthimiopoulos}. 

Our present interest in the emission spectra of xenon-doped liquid argon comes from a recent discovery of a strong emission from this medium in the infrared (IR) at $\sim1.18\mu$m peak wavelength \cite{Neumeier_IR}. Since two parallel optical signals could improve the discrimination power of an event-by-event particle identification in a scintillation detector, we studied the corresponding emission spectrum in the VUV for the xenon concentration of 10\,ppm. For this concentration the infrared emission has maximum light output \cite{Neumeier_IR}. If it turns out that the intensity ratio or the timing signals, or both, for the IR and VUV depend on the type of the incident particle, it will be possible to build a particle detector in which different particles can be identified on a pure optical basis (IR and VUV detector for the scintillation light).

\section{Experimental Setup}

The excitation technique of the liquid target was identical to that of previous IR experiments \cite{Neumeier_IR}. Briefly, a 12\,keV electron beam was sent through a 300\,nm thin silicon-nitride/silicon-oxide membrane into the liquid. Light was detected via an elliptical imaging optics and a f=30\,cm monochromator (McPherson model 218) in combination with a phototube with MgF$_{2}$ entrance window and S20 cathode (ET Enterprises, D860B) operated in photon counting mode replacing the InAs IR detector used in ref.\,\cite{Neumeier_IR}. The gas mixture was prepared at room temperature, purified (SAES getters, Model: MONO TORR$^{\textrm{\textregistered}}$ PS4-MT3-R-2) and then condensed in the target cell. A metal bellows pump kept the gas in a continuous flow in a closed cycle where the gas was evaporated, purified and condensed into the cell again via a heat exchanger. A heating resistor was used to maintain a constant temperature ($\sim86$\,K) and the flow of gas through the cryogenic cell was adjusted via a bypass valve. A more detailed description of the experimental setup can be found in ref.\,\cite{Heindl_2}.

\section{Spectral Shape and Scintillation Efficiencies of Pure and Xenon-Doped Liquid Argon}
  \label{sec:results}
\begin{figure}
  \centering
  \includegraphics[width=\columnwidth]{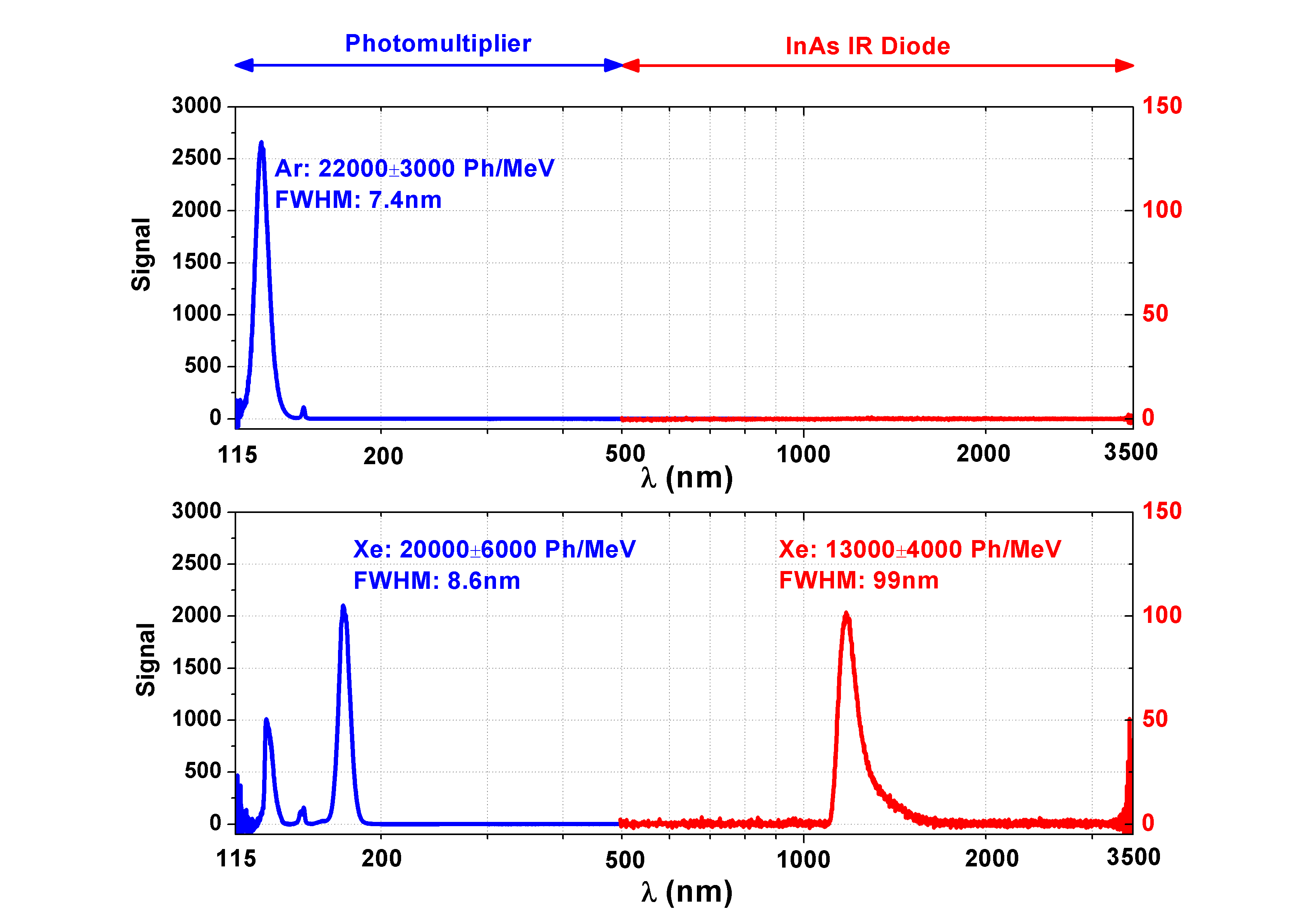}
  \caption{\textit{Electron-beam induced VUV and IR emission spectra of pure liquid argon (upper panel) and liquid argon doped with 10\,ppm xenon (lower panel). In the VUV, the Xe doping leads to a shift of the signal from the 127\,nm Ar emission to the 174\,nm xenon emission. In addition to the VUV, an IR emission is formed at $\sim1.18\,\mu$m peak wavelength (see ref.\,\cite{Neumeier_IR} for details). Note the wide wavelength range in a logarithmic representation from 115\,nm to 3.5\,$\mu m$ and the differently scaled ordinates for the VUV and the IR parts of the measurement. The scaling of the ordinates represents the measured integral efficiencies of the main emissions in photons/MeV (see text for details). The emission spectra from 115\,nm to 500\,nm were measured with a photomultiplier with S20 cathode operated in photon counting mode. The wavelength range from 500\,nm to 3.5\,$\mu m$ was scanned with an InAs infrared photodiode \cite{Neumeier_IR}. The small peak at 149\,nm in the upper panel is caused by a residual xenon impurity which could not be removed completely.}}
  \label{fig:VUV_IR_Uebersicht}
\end{figure}

\begin{figure}
 \centering
 \includegraphics[width=\columnwidth]{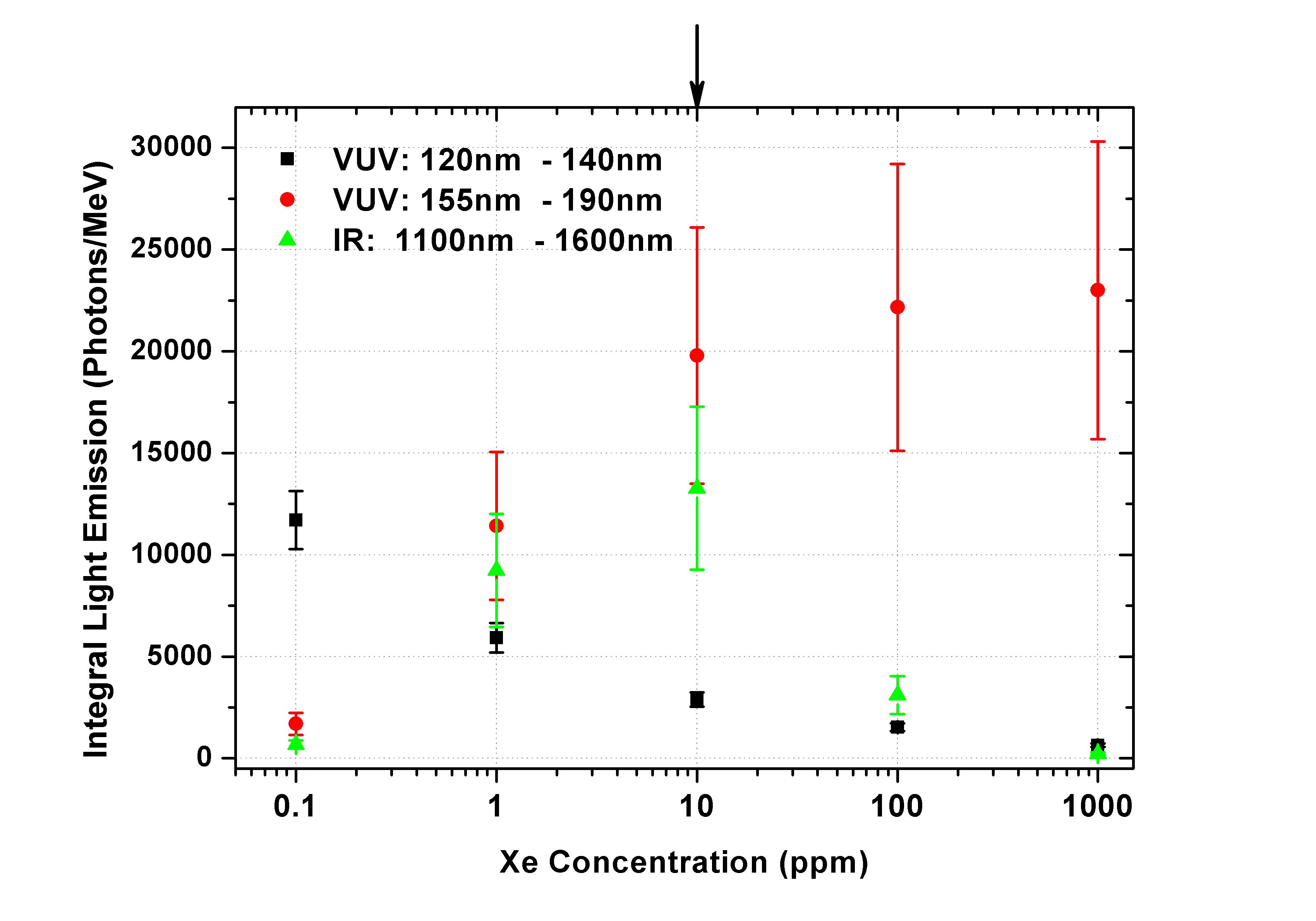}
 \caption{\textit{Integral amount of light in photons/MeV from the argon VUV emission (120\,-\,140\,nm, black squares) and the xenon VUV emission (155\,-\,190\,nm, red dots) for different xenon concentrations in liquid argon. The green triangles represent the integral amount of light in photons per MeV from the IR emission (1100\,nm - 1600\,nm) of the different mixtures. The energy transfer from argon to xenon is clearly demonstrated by the decreasing argon emission and the increasing xenon emission. At 10\,ppm the energy transfer in the VUV is almost complete (86\% compared to the 1000\,ppm mixture), indicated by the  arrow in the upper part of the figure. This mixture also shows maximum light output in the IR region \cite{Neumeier_IR}.}}
 \label{fig:VUV_Emission_Integral}
\end{figure}

The main question in the context of the motivation described above is the shape of the emission spectrum and the scintillation efficiency of liquid argon doped with 10\,ppm xenon in a wide wavelength range from the VUV to the IR. With the two different detectors (VUV phototube in the present work and InAs photodiode in ref.\, \cite{Neumeier_IR}) attached to the monochromator we could record the electron-beam induced light emission from pure and xenon-doped liquid argon from 115\,nm to 3.5\,$\mu m$. The result is presented in Fig.\,\ref{fig:VUV_IR_Uebersicht}. A logarithmic wavelength axis is used to show the wide wavelength range in a reasonable way. The efficiency values presented in Fig.\,\ref{fig:VUV_IR_Uebersicht} were determined by a wavelength integrated measurement in the regions of the main emission features (see Fig.\,\ref{fig:VUV_Emission_Integral} and the discussion below). 

It has to be noted that in the VUV and in the IR region the relative spectral detector responsivities have been included in the representation of the data. Furthermore it is important to distinguish between the power of the emitted light and the number of photons. The ratio between the photon energy at 174\,nm and $\sim$1.18\,$\mu$m is $\sim6.8$.
The number of photons emitted in the ranges from 115 to 145\,nm (upper panel, pure liquid argon) and from 155 to 190\,nm (lower panel, liquid argon doped with 10\,ppm xenon) per MeV electron-beam power deposited was determined in the following way: The cell was filled with pure argon and xenon, respectively, in the gas phase at 1\,bar and room temperature. A spectrum was recorded with an electron beam current of 1.3\,$\mu$A ($(11.1 \pm 1.5)$\,mW for pure argon and $(9.1\pm 1.3)$\,mW for pure xenon with the energy loss in the entrance foil and the backscattered electrons taken into account\footnote{The mean energy deposited in the gas by a 12\,keV electron has been obtained by simulating the transmission of the 300\,nm thin silicon-nitride/silicon-oxide membrane using GEANT4 \cite{GEANT4}}). The conversion efficiency from electron-beam power to excimer-light ($(33 \pm 4)\%$ for pure gaseous argon and $(42 \pm 5)\%$ for pure gaseous xenon) had been carefully measured in an analogue setup with a calibrated VUV detector also using a 12\,keV electron beam \cite{Morozov_Effizienz}. Then, pure liquid argon and liquid argon doped with 10\,ppm xenon, respectively, were prepared and condensed into the cell. The only setting which was changed from the gaseous to the liquid measurements was the steering of the electron beam and the alignment of the light collecting mirror because of a slight change of the position of the target cell due to the thermal contraction of its holder during cooling. Because of the practically identical spectral shape of the light emission from pure liquid argon compared to gaseous argon and from the liquid argon-xenon mixture compared to gaseous xenon, the efficiency for the liquids could be determined as $(21 \pm 3)\,\%$ energy efficiency, corresponding to $22000 \pm 3000$\,photons/MeV (120\,nm - 140\,nm) for pure liquid argon and $(14 \pm 4)\%$\footnote{The influence on the efficiency of a different amount of backscattered electrons in the transfer from gaseous xenon to liquid argon doped with 10\,ppm xenon has been taken into account using a GEANT4 \cite{GEANT4} simulation (accuracy: 10\,\%). The mean energy deposited by a 12\,keV electron sent through the silicon-nitride/silicon-oxide membran into liquid argon doped with 10\,ppm xenon is 8.5\,keV. When the cell is filled with pure gaseous xenon, the mean energy is 7.0\,keV.} energy efficiency, corresponding to $20000 \pm 6000$ photons/MeV (155\,nm - 190\,nm) for liquid argon doped with 10\,ppm xenon, respectively.

A corresponding value of $13000 \pm 4000$ photons/MeV according to $(1.3 \pm 0.4)\%$ energy efficiency had been found for the infrared emission. This value and the corresponding error margins could be obtained from a detailed analysis of the systematic errors of the experiment described in ref.\,\cite{Neumeier_IR}. In summary, a liquid mixture of 10\,ppm xenon in argon is a scintillator which provides a comparable signal in terms of photons/MeV in both the VUV and the IR spectral range.

The authors are aware of the fact that in the case of pure liquid argon the measured scintillation efficiency ($22000 \pm 3000$\,photons/MeV) is approximately a factor of two below the efficiency of $40000$\,photons/MeV measured by Doke et al. \cite{Doke}. Note, however, that Doke et al. used electrons with an energy of 0.976\,MeV from a radioactive source\footnote{Internal conversion electrons from $^{207}$Bi.} compared to 12\,keV ($\sim 8.5$\,keV after the entrance membrane) electrons from a cathode ray tube in this experiment. These low-energy electrons used here lose much more energy per unit length than the relativistic electrons of Doke et al. which could lead to a decreased scintillation efficiency due to the scintillator non-proportionality which is caused by exciton-exciton quenching \cite{Scintillator_Nonproportionality}. To the best of our knowledge there exists no other published measurement of the scintillation efficiency of pure liquid argon using such low-energy electrons. A variation of the electron beam current from 1.3$\mu$A down to 100\,nA showed a constant scintillation efficiency. However, track-to-track quenching caused by too high a beam current cannot be excluded since the beam current could not be reduced below 100\,nA in a controlled way.

A GEANT4 simulation showed that the mean track length the electrons travel before they are stopped is approximately 2.7\,$\mu$m in liquid argon. The mean penetration depth of the electrons in the direction perpendicular to the entrance membrane is approximately 1.5\,$\mu$m. Therefore, the excitation happens in a very thin layer of argon close to the entrance membrane. A radiationless deexcitation of atoms and molecules at the surface of the membrane could also explain the reduced scintillation efficiency compared to \cite{Doke}. 

It has to be emphasized that the efficiency values stated here are derived from the absolute measurements in pure gaseous argon and pure gaseous xenon published in \cite{Morozov_Effizienz}. There, a NIST/PTB calibrated optical semiconductor detector (SXUV-100, IRD inc.) was used to measure the excimer-light intensities (without any optical windows) for exactly the same excitation method as in the experiments presented here. Therefore, the measured scintillation efficiencies in the VUV for pure liquid argon as well as liquid argon doped with xenon can be traced back to a measurement with a calibrated optical detector.  

\begin{figure}
  \centering
  \includegraphics[width=\columnwidth]{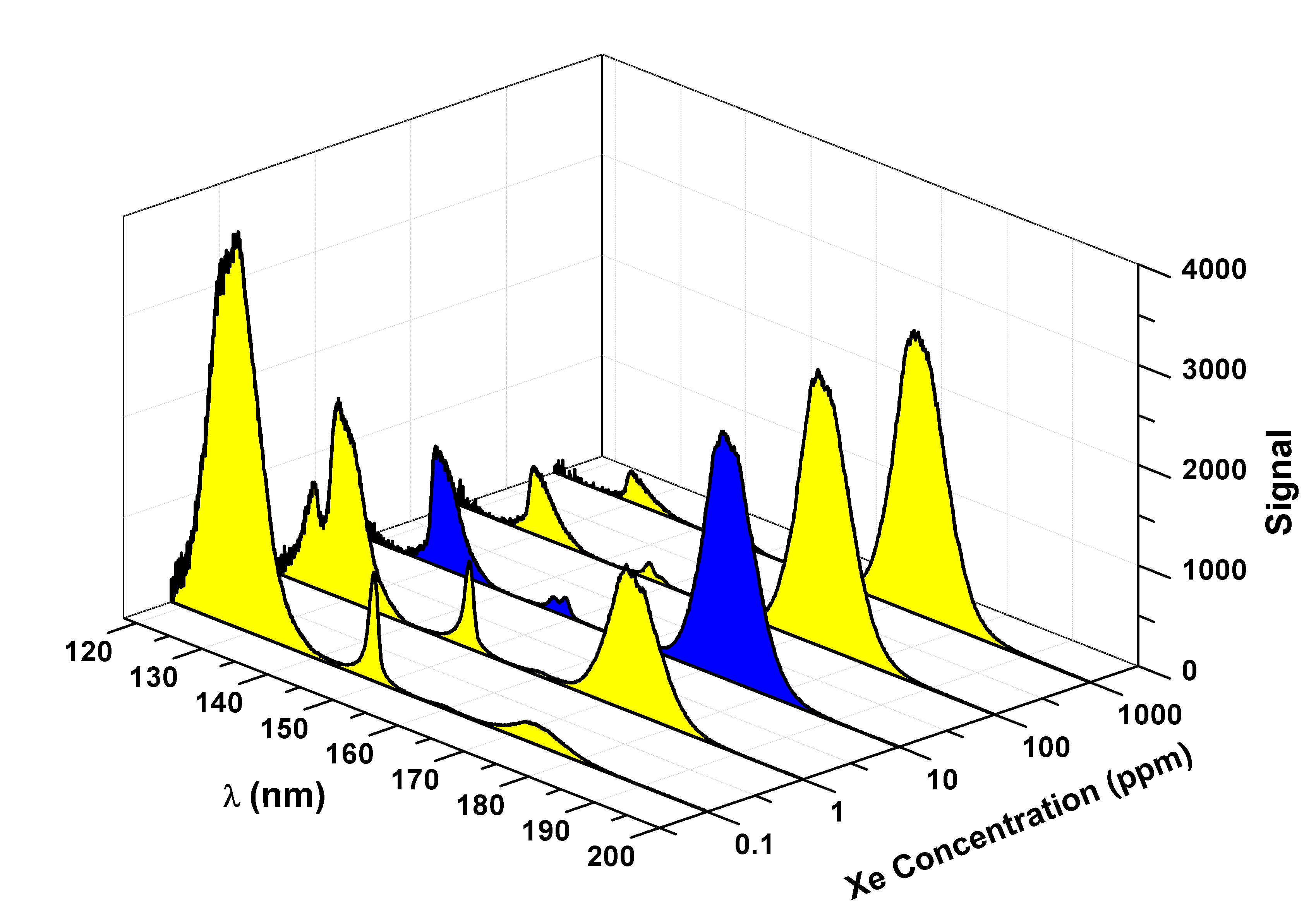}
  \caption{\textit{Electron-beam induced VUV emission spectra of mixtures of 0.1, 1, 10, 100 and 1000\,ppm xenon in liquid argon. An energy transfer from the argon emission at 127\,nm to the xenon emission at 174\,nm with increasing xenon concentration is clearly visible. At 10\,ppm xenon in argon the energy transfer in the VUV is almost complete and this mixture has a maximum in the light output in the IR region. The spectra were corrected applying the detector response function. For the explanation of the absorption dip at $\sim130$\,nm, see Fig.\,\ref{fig:VUV_Emission_Transmission}.}}
  \label{fig:VUV_Emission_3D}
  \end{figure}

\begin{figure}
    \centering
    \includegraphics[width=\columnwidth]{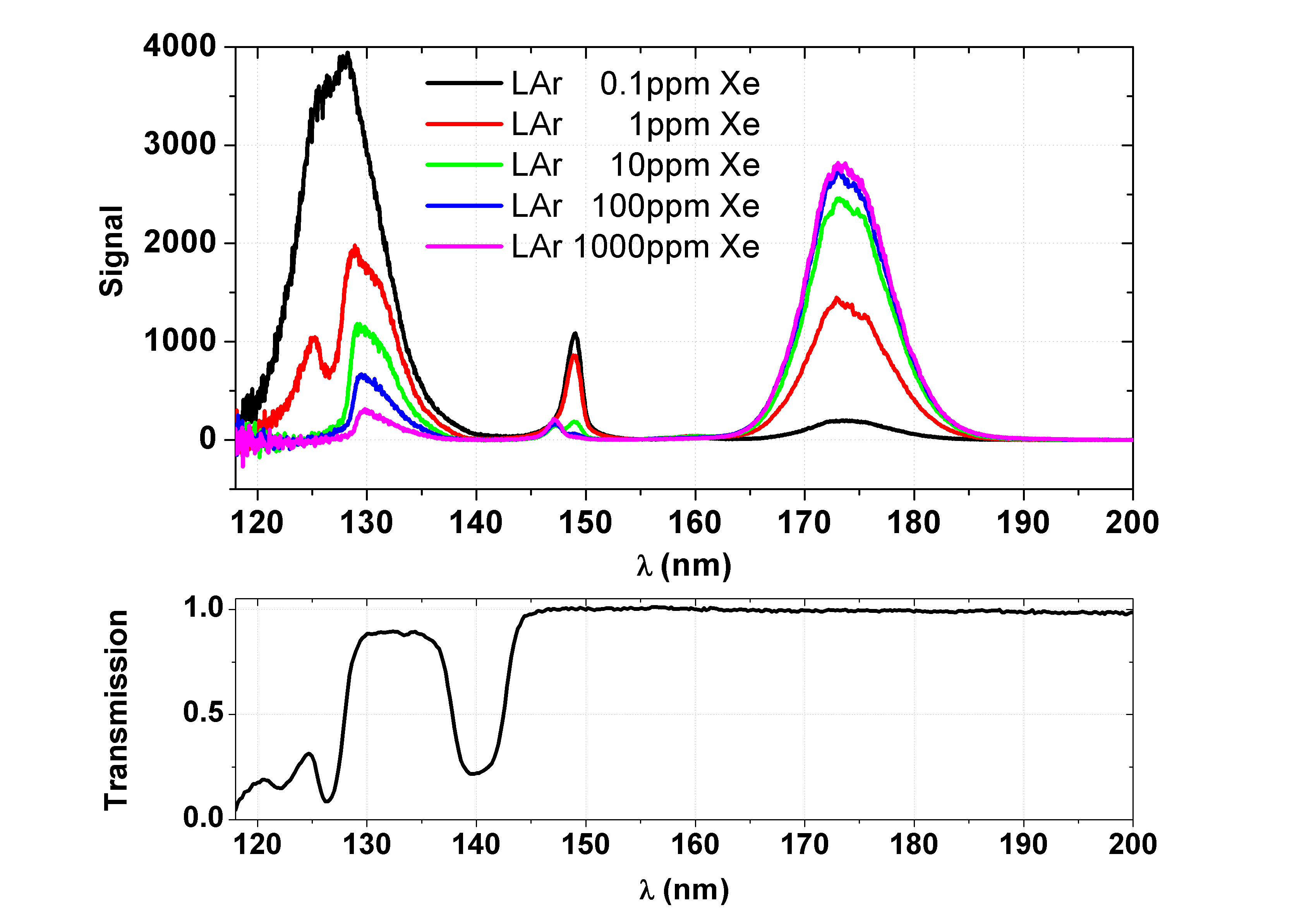}
    \caption{\textit{The VUV emission of electron-beam excited liquid argon doped with 0.1, 1, 10, 100, 1000\,ppm xenon is shown (upper panel). A decreasing argon-excimer emission at 127\,nm peak wavelength and an increasing xenon-excimer emission at 174\,nm peak wavelength are visible. The spectra were corrected applying the detector response function. However, below 140\,nm the spectra are overestimated by a factor of approximately two due to a systematic error in the detector response in that wavelength region. Note that at 10\,ppm xenon in liquid argon the energy transfer is almost complete (see also Fig.\,\ref{fig:VUV_Emission_Integral}). The 10\,ppm mixture has maximum light output in the IR region. The lower panel shows the transmission of liquid argon doped with 0.1\,ppm xenon measured with a deuterium light source and a length of the optical path of 116\,mm in a not yet published experiment. In the experiments described here the light had to travel only 2\,mm from the light emitting spot at the entrance membrane to the first optical window.}}
    \label{fig:VUV_Emission_Transmission}
 \end{figure}

The energy transfer mechanism leading to the strong change of the emission spectrum of pure liquid argon (ref. \cite{Heindl_1,Heindl_2} and Fig.\,\ref{fig:VUV_IR_Uebersicht} upper panel) to the one of the mixture with the tiny amount of xenon (Fig.\,\ref{fig:VUV_IR_Uebersicht} lower panel) will be studied in detail spectroscopically including time-resolved measurements. The modification of the VUV emission spectrum by adding xenon to argon is displayed in Figs.\,\ref{fig:VUV_Emission_3D} and \ref{fig:VUV_Emission_Transmission} (upper panel) in the form of wavelength spectra recorded with xenon admixtures ranging from 0.1\,ppm to 1000\,ppm. Fig.\,\ref{fig:VUV_Emission_Integral} shows the integral intensity values of light emission with increasing xenon concentration for the VUV as well as the IR emission features. The data points in Fig.\,\ref{fig:VUV_Emission_Integral} were obtained by integrating the raw data spectra from Fig.\,\ref{fig:VUV_Emission_3D} for the VUV and Fig.\,2 in ref.\,\cite{Neumeier_IR} for the IR and normalizing them to the measured efficiencies in photons/MeV.

The shift of the main VUV emission from 127\,nm to 174\,nm is a benefit for scintillation detectors in two ways: Firstly, the VUV emission is shifted to a wavelength region where photomultipliers generally become more sensitive due to the spectral responsivities of the cathode materials and the increased transmissions of the window materials. One could also think of using photomultipliers without wavelength shifters and quartz windows. Secondly, the main emission is shifted to a wavelength region where a high transmission of xenon-doped liquid argon has been measured (see Fig.\,\ref{fig:VUV_Emission_Transmission}, lower panel). Xenon in liquid argon leads to a transmission minimum at 127\,nm. Therefore xenon can be a critical impurity, in particular for large liquid-argon detectors. A detailed analysis explaining the origin of the transmission minima can be found in ref.\,\cite{Neumeier} and references therein. A detailed study of the absorption of VUV light for the 0.1\,ppm to 1000\,ppm mixtures will be published in a separate paper.

For the scintillation-detector aspect it is important to note that the energy transfer from argon to xenon begins to saturate already for the 10\,ppm mixture. A comparison of the data in Fig.\,\ref{fig:VUV_Emission_Integral} with those presented in Fig.\,7 of reference \cite{Efthimiopoulos} shows that in the liquid at 0.1\,ppm the ratio of the integral VUV emission features related to argon and xenon, respectively, correspond roughly to those at 100\,ppm in the gas phase. An emission of the first excimer continuum of xenon also appears in the liquid phase but can only be seen in logarithmic representation of the data. It is also interesting to note that infrared lasers in dense, gaseous Ar-Xe mixtures have optimum operation conditions around 3000\,ppm to 5000\,ppm Xe in Ar \cite{Ulrich_e_beam_laser,Skrobol_e_beam_laser}. The VUV emission spectra are dominated by the 147\,nm Xe line \cite{Sventitskii} under these conditions.

\section{Time Structure of the VUV and the NIR Emission in Xenon-Doped Liquid Argon}
\begin{figure}
 \centering
 \includegraphics[width=\columnwidth]{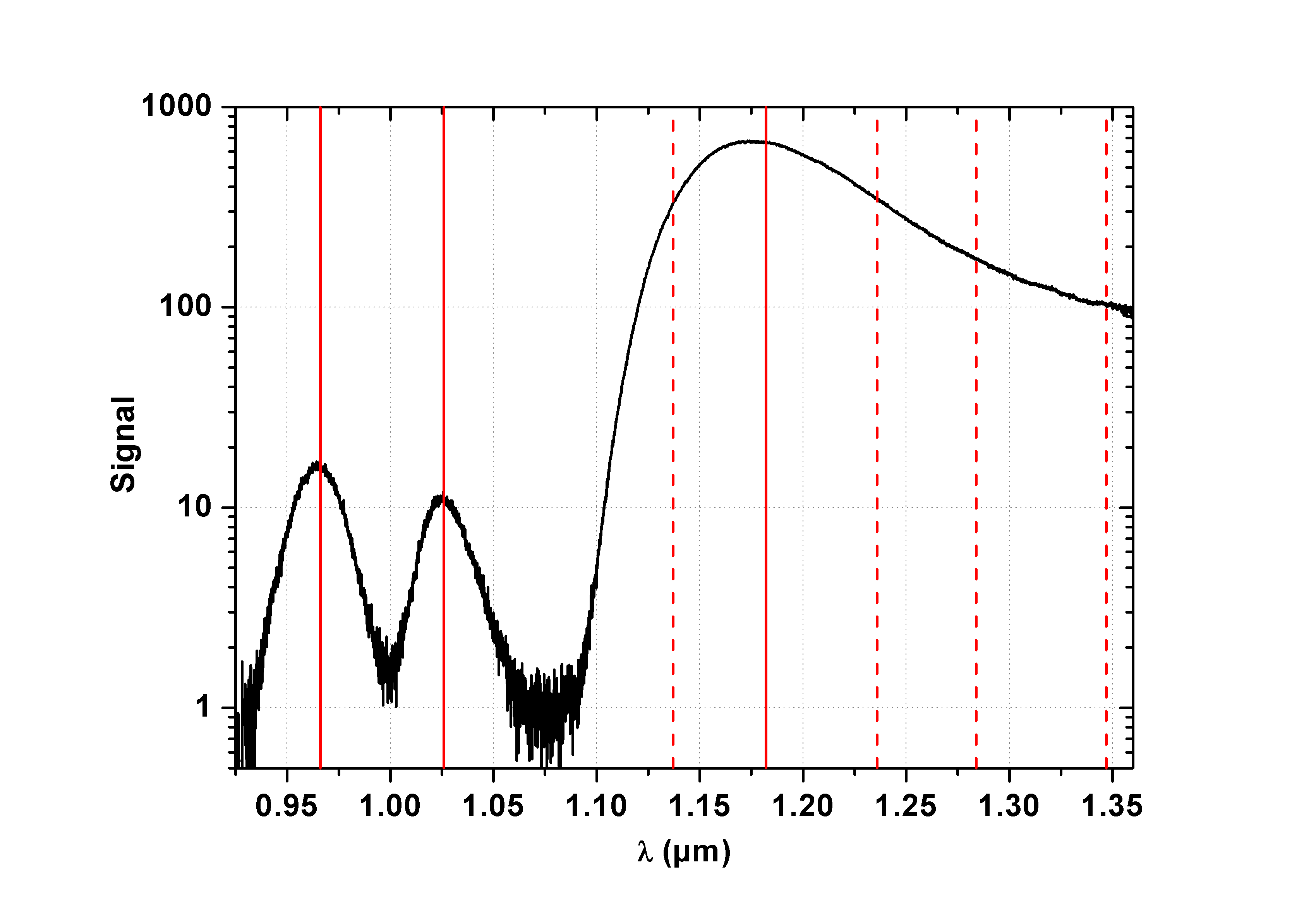}
 \caption{\textit{A near-infrared emission spectrum of electron-beam induced scintillation of liquid argon doped with 10\,ppm xenon is shown. The spectrum has been measured with a near-infrared photomultiplier (Hamamatsu NIR-PMT Module H10330B-45 SEL) from 925\,nm to 1360\,nm with a wavelength resolution of 0.5\,nm (slit width: 50\,$\mu$m). The signal has been corrected applying the detector responsivity provided by the manufacturer. The intense, asymetrically shaped $\sim1.18\mu$m IR emission discovered in ref.\,\cite{Neumeier_IR} could be confirmed and two new minor symmetrically shaped emission features centered at 964\,nm and 1024\,nm, respectively were discovered (see text for details). The solid red lines show the wavelength positions of the time structures presented in Figs.\,\ref{fig:Zeitstruktur_VUV_IR_Vergleich} and \ref{fig:Zeitstruktur_IR_Nebenmaxima}, respectively. The red dashed lines denote wavelength positions where additional time structure measurements have been performed.}}
 \label{fig:Hamamatsu_IR_Emission}
\end{figure}

To judge whether the newly found VUV and IR emissions could be used in a particle detector the time structure has been measured using a standard time to amplitude conversion technique. The electron beam was pulsed with a repetition rate of 8\,kHz and the duration of the pulses was $\sim$100\,ns. The time structure of the IR emission was measured with a near-infrared photomuliplier (Hamamatsu NIR-PMT Module H10330B-45 SEL) attached to the monochromator instead of the InAs photodiode. Due to the increased sensitivity and the huge dynamic range of the photomultplier compared to the InAs photodiode it was possible to measure the IR emission with increased resolution. In Fig.\,\ref{fig:Hamamatsu_IR_Emission} the electron-beam induced emission spectrum of liquid argon doped with 10\,ppm xenon measured with the NIR-PMT from 925\,nm to 1360\,nm is presented. The spectrum shows two new symmetrically shaped emission features centered at 964\,nm (FWHM:\,28\,nm) and at 1024\,nm (FWHM:\,27\,nm), respectively. Until now, these emissions features could not be resolved using the InAs photodiode. The emission feature centered at 964\,nm was probably already observed in pure liquid argon by Heindl et al. (see Fig.\,2 in ref.\,\cite{Heindl_1}). In our last publication \cite{Neumeier_IR} we believed that this emission was an artifact due to the normalization of the spectrum with the response function of the spectrometer used for that spectral region. A measurement of the emission spectrum of pure liquid argon with the NIR-PMT will finally clarify this issue. The intense asymetrically shaped IR emission peaking at $\sim$1.18$\mu$m \cite{Neumeier_IR} could be confirmed and the exact peak wavelength is at 1173\,nm (FWHM:\,99\,nm).

\begin{figure}
 \centering
 \includegraphics[width=\columnwidth]{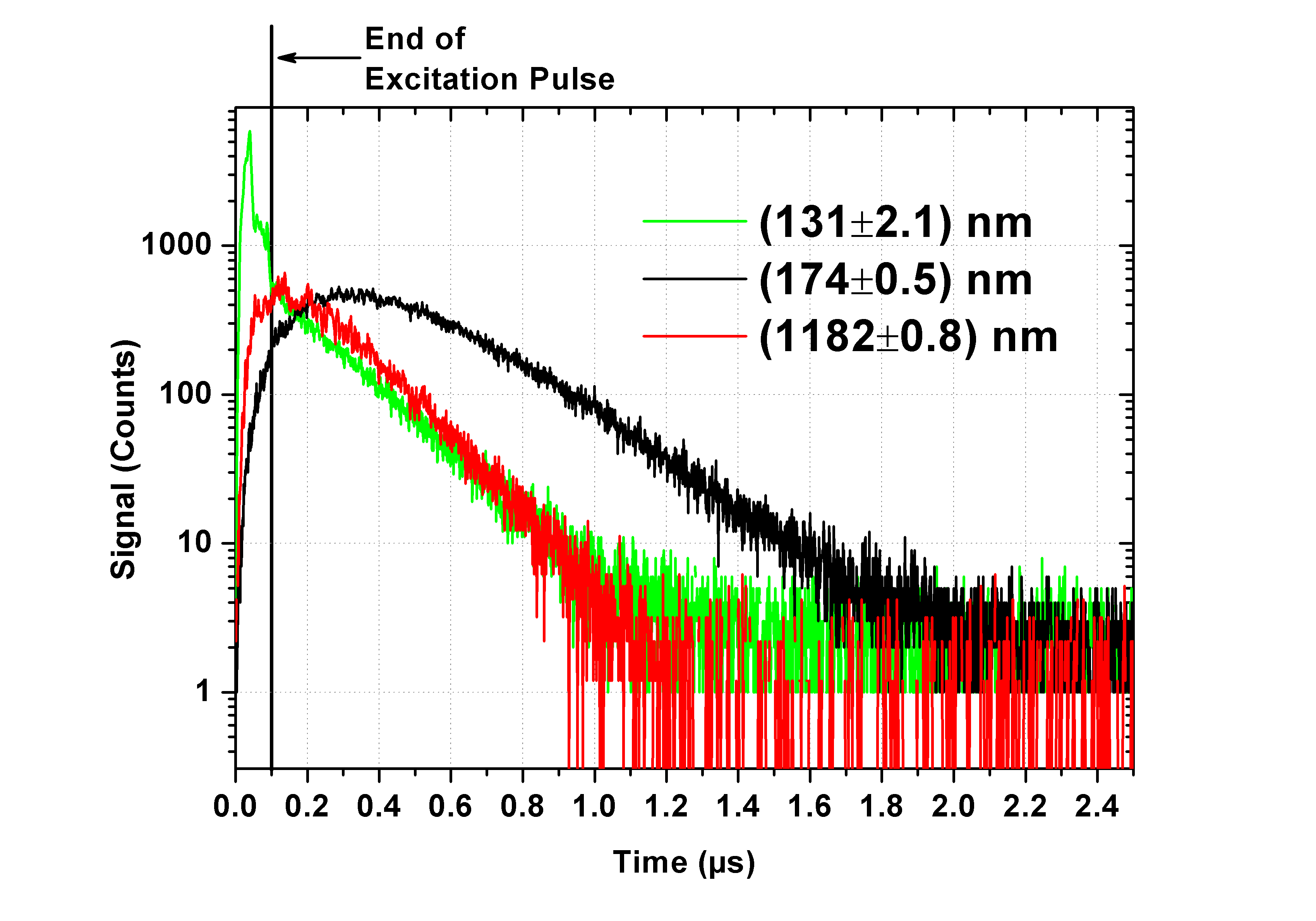}
 \caption{\textit{The time structures of the electron-beam induced scintillation (VUV and NIR) of liquid argon doped with 10\,ppm xenon are shown. The excitation pulse had a duration of $\sim100$\,ns with a sub-structure in intensity. The black curve shows the time structure of the most intense emission feature in the VUV at 174\,nm (Xe excimer). The red curve shows the time structure of the most intense emission in the near-infrared part of the emission spectrum measured at a wavelength of 1182\,nm. The emission in the VUV at 131\,nm is also shown for comparison. It can be attributed to argon excimers (see text for details). The decay-time constants are: $(202\pm2)$\,ns (fitted from $0.2\mu$s to $1\mu$s) for the emission at 131\,nm, $(240\pm10)$\,ns (fitted from $1.0\mu$s to $1.6\mu$s) for the emission at 174\,nm and $(173\pm3)$\,ns (fitted from $0.4\mu$s to $0.8\mu$s) for the emission at 1182\,nm.}}
 \label{fig:Zeitstruktur_VUV_IR_Vergleich}
\end{figure}

Fig.\,\ref{fig:Zeitstruktur_VUV_IR_Vergleich} presents the measured time structures of liquid argon doped with 10\,ppm xenon at 131\,nm, 174\,nm and 1182\,nm where the latter two corresspond to the most intense emission wavelengths of the mixture. The data are shifted in time to match at zero. However, the VUV measurements at 131\,nm and at 174\,nm have the same temporal zero point since these time structures were measured using the same photomultiplier (VUV-PMT). The NIR-PMT has a different internal delay compared to the VUV-PMT. A beam-current variation by a factor of 16 did not change the measured time structures at 174\,nm as well as at 1182\,nm. Furthermore, a variation of the photon rate did not have an influence on the measured time structures. In the measurements the rate was adjusted by the slits of the monochromator so that approximately one photon was detected every 50th to 100th pulse. The time range of the time-to-amplitude converter was calibrated using a precision pulse generator (Stanford Research DG535).  

The VUV emission at 174\,nm peaks at $\sim250$\,ns, i.e., $\sim150$\,ns after the end of the excitation pulse. After that the light intensity decreases. After  $\sim1\mu$s the time structure can be described by an exponential decay with a decay time of $(240\pm10)$\,ns. However, so far we hesitate to present a model describing the time structures observed. Such a model would require concentration-dependent measurements of the time structures to identify the different processes and the corresponding rate constants of the reaction chains. Therefore all the decay-time constants presented in the following are just phenomenological best fit values of exponential decays which are fitted to the data. 

The time structure of the IR main emission feature has been measured at five different wavelength positions (1137, 1182, 1236, 1284, 1347\,nm) indicated by the red vertical lines in Fig.\,\ref{fig:Hamamatsu_IR_Emission}. The decay-time constants at these five different wavelengths had a maximum deviation of 15\% around a mean value of $175$ns. This is a strong hint that in this broad IR emission feature no further IR emitting species with significantly different time constants are involved. Time-structure measurements at more wavelength positions with much higher statistics have to be performed to reduce the statistical errors and to clarify whether a wavelength dependent time structure is hidden within this intense IR emission. Therefore, the discussion below is related only to the measured time structure at the peak of the emission at 1182\,nm.

The IR emission at 1182\,nm peaks faster than the 174\,nm VUV emission. The decay of the 174\,nm VUV emission can well be described by a single exponential decay only after the IR emission has decayed completely. In the time interval where the 1182\,nm emission is still decaying the time constant of the 174\,nm emission changes from a slower to a faster decay. This could be a hint that the 174\,nm VUV emission is a transition in a cascade following the 1182\,nm emission. 

The VUV emission at 131\,nm (compare Fig.\,\ref{fig:VUV_Emission_3D}) shows a fast component basically reproducing the excitation pulse and a slow component with a time constant of $(202\pm2)$\,ns (see Fig.\,\ref{fig:Zeitstruktur_VUV_IR_Vergleich}). This feature can be attributed to the emission of argon excimers where the triplet liftime is strongly reduced due to the energy transfer from argon to xenon. In pure liquid argon the triplet lifetime is of the order of a microsecond \cite{Heindl_1}.
In conclusion, the two main emission bands measured at 174\,nm and at 1182\,nm have decayed completely after approximately 2.0\,$\mu$s and 1.2\,$\mu$s, respectively. These time structures are well suited for a particle detector, since no long-living components are involved which would increase the noise level and the dead-time of a detector. 

\begin{figure}
 \centering
 \includegraphics[width=\columnwidth]{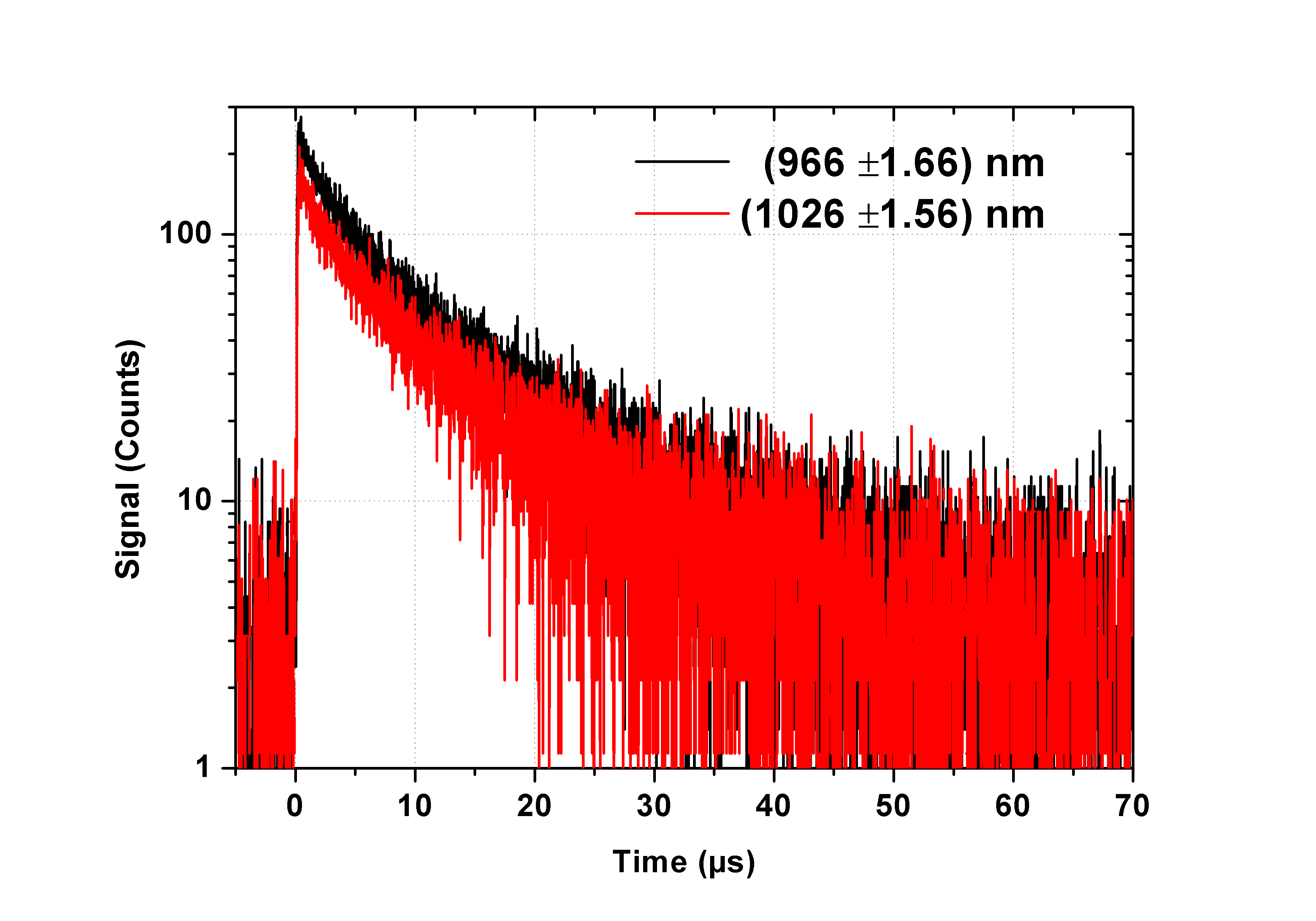}
 \caption{\textit{The time structures of the electron-beam induced scintillation of the newly found near-infrared emission features are presented. The excitation pulse had a duration of approximately 100\,ns. The black curve shows the time structure at a wavelength of 966\,nm and the red curve at a wavelength of 1026\,nm. The decay times are much longer than the decay times of the main emission features (174\,nm, black curve in Fig.\,\ref{fig:Zeitstruktur_VUV_IR_Vergleich} and 1182\,nm, red curve in Fig.\,\ref{fig:Zeitstruktur_VUV_IR_Vergleich}).}}
 \label{fig:Zeitstruktur_IR_Nebenmaxima}
\end{figure}

The time structures of the newly discovered emissions at 964\,nm and 1024\,nm peak wavelength have also been measured and the time spectra are presented in Fig.\,\ref{fig:Zeitstruktur_IR_Nebenmaxima}. These emissions can be compared directly since they were measured with the same NIR-PMT. 

\section{Summary and Outlook}
In summary, it has been found that the efficient energy transfer from argon to xenon which is well known from the gas phase is also present in the liquid phase. This had already been observed in the context of recording emission spectra from nominally pure liquid argon \cite{Heindl_2,Heindl_1,Hofmann} and in Fig.\ref{fig:VUV_IR_Uebersicht} (upper panel) where a small residual emission of xenon is visible at 147\,nm indicating an impurity. The important result of the present study is that 10\,ppm xenon in liquid argon are sufficient to shift the main emission in the VUV from the argon-excimer emission (127\,nm peak wavelength) almost completely to the xenon-excimer emission (174\,nm peak wavelength). An additional infrared emission at 1.17$\mu$m peaks at this xenon concentration.

The time structures of both emissions are well suited for a particle detector since the emissions have short decay times which leads to a low noise level and a low dead-time in a particle detector. It seems that a particle identification based on the singulet to triplet ratio like in pure liquid argon detectors is not possible anymore (see e.g. refs.\,\cite{DEAP_1,DEAP_2}). However, a second intense signal in the near-infrared may have the potential of developing scintillation detectors with simultaneous light detection in the VUV and IR for particle identification. Experiments to study this aspect are in preparation.

\section*{Acknowledgements}
This research was supported by the DFG cluster of excellence "Origin and Structure of the Universe" (www.universe-cluster.de) and by the Maier-Leibnitz-Laboratorium in Garching.


\begin{thebibliography}{99}
\bibliographystyle{unsrt}

  \bibitem{ArDM}
	ArDM Collaboration,\\
	arXiv:1307.0117v1[physics.ins-det] (2013)
    
  \bibitem{DarkSide}
        A. Wright (DarkSide Collaboration),\\
        arXiv:1109.2979v1[physics.ins-det] (2011)
        
  \bibitem{MiniCLEAN}
        K.\,Rielage et al.,
        arXiv:1403.4842v1[physics.ins-det] (2014)
        
  \bibitem{DEAP_1}
        M. G. Boulay, 
        J. Phys.: Conf. Ser. \textbf {375}, 012027 (2012)
        
  \bibitem{DEAP_2}
	DEAP Collaboration,\\
	arXiv:1406.0462v1[astro-ph.IM] (2014)
  
  \bibitem{DARWIN}
        M. Schumann (DARWIN consortium),\\
        arXiv:1111.6251v1[astro-ph.IM] (2011)
        
  \bibitem{ICARUS}
	M.\,Antonello et al.,\\
	arXiv:1312.7252v2[physics.ins-det] (2014)
  
  \bibitem{LBNE}
	LBNE Collaboration
	arXiv:1307.7335v3[hep-ex] (2014)
  
  \bibitem{Pollmann}
        P. Peiffer  et al.,
        JINST \textbf{3}, P08007 (2008)
        
  \bibitem{Excimer_Buch}
        Ch. K. Rhodes \and C.A. Brau, Excimer Lasers, Springer-Verlag, Berlin (1984)
        
  \bibitem{Efthimiopoulos}
        T. Efthimiopoulos  et al., 
        J. Phys. D \textbf{30} (1997)
        
  \bibitem{Neumeier_IR}
        A. Neumeier et al.,
        EPL \textbf{106}, 32001 (2014)
        
  \bibitem{Heindl_2}
        T.\,Heindl et al.,
        JINST \textbf {6}, P02011 (2011)
        
  \bibitem{Morozov_Effizienz}
        A. Morozov  et al.,
        J. Appl. Phys.\textbf{103}, 103301 (2008)
        
  \bibitem{GEANT4}
        Geant4 Collaboration,\\
        Nucl. Instrum. Methods A \textbf{506}, 250 (2003)   
     
  \bibitem{Doke}
        T. Doke et al.,
        Nucl. Instrum. Methods A \textbf{291}, 617 (1990)
  
  \bibitem{Scintillator_Nonproportionality}
        S. Payne et al.
        IEEE Trans. Nucl. Sci. \textbf{58}, 6 (2011) 
     
  \bibitem{Heindl_1}
        T.\,Heindl et al.,
        EPL \textbf {91}, 62002 (2010)   
   
  \bibitem{Neumeier}
        A. Neumeier et al.,
        Eur. Phys. J. C, \textbf{72}, 2190 (2012)
        
  \bibitem{Ulrich_e_beam_laser}
        A. Ulrich  et al., 
        J. Appl. Phys. \textbf{86}, 3525 (1999) 
        
  \bibitem{Skrobol_e_beam_laser}
        C. Skrobol  et al.,
        Eur. Phys. J. D \textbf{54}, 103 (2009)      
        
  \bibitem{Sventitskii}
        A.R. Striganov \and N.S. Sventitskii, 
        \textit{TABELS OF SPECTRAL LINES OF NEUTRAL AND IONIZED ATOMS}, Vol. 9, IFI/Plenum Data Corporation New York, Washington, p. 572 (1968)     
           
  \bibitem{Hofmann}
        M. Hofmann  et al.,
        Eur. Phys. J. C \textbf{73}, 2618 (2013)
\end{thebibliography}
\end{document}